%% file: manuscriptrev_v2.tex
\documentclass[aps,prd,twocolumn,showpacs,superscriptaddress,nofootinbib,groupedaddress]{revtex4-1} 
\usepackage{graphicx}
\usepackage{epsfig}
\usepackage{tabularx}
\usepackage{booktabs}
\usepackage{amssymb}
\usepackage{amsmath}
\usepackage{natbib}
\RequirePackage[colorlinks,citecolor=blue,urlcolor=blue,linkcolor=blue]{hyperref}
\usepackage{dcolumn}
\begin{document} 
\preprint{PRD}

%Title of paper
\title{Modulated bimaximal neutrino mixing}

\author{S.~Roy}
 \email{meetsubhankar@gmail.com}
\affiliation{Department of Physics, Gauhati University, Guwahati, Assam -781014, India
}%
\author{N.~N.~Singh}
 \email{nimai03@yahoo.com}
\affiliation{Department of Physics, Manipur University, Imphal, Manipur-795003, India
}%
\date{\today}

%\date{\today}

%====================================
 
\begin{abstract}
The present article is an endeavor to look into some fruitful frameworks based on ``Bi-maximal'' neutrino mixing, from a model independent stand. The possibilities involving the correction or attenuation of the original BM mixing matrix, followed by GUT-inspired charged lepton correction are invoked. The ``symmetry-basis'' thus constructed, accentuates some interesting facets such as: a modified QLC relation, $\theta_{12}+\theta_{c}\approx\frac{\pi}{4}-\theta_{13}\cos(n\pi-\delta_{CP})$, a possible link up between neutrino and charged lepton sectors, $\theta_{13}^{\nu}=\theta_{12}^{l}\sim\mathcal{O}(\theta_{C})$ or that between neutrinos and quarks, $\theta_{13}^{\nu}=\theta_{C}$. The study vindicates the relevance of the Bi-maximal mixing as a first approximation.
\end{abstract}

%=====================================
\pacs{11.30.Hv 14.60.-z 14.60.Pq 14.80.Cp 23.40.Bw}
\keywords{Neutrino masses, Neutrino mixing, Cabibbo angle, $\mu$-$\tau$ symmetry}
\maketitle
\section{Introduction}
In the past few years, certain significant description are imprinted by the oscillation experiments\,\cite{Abe:2013hdq, An:2013zwz, Adamson:2013whj, Ahn:2012nd}, which compels one to rethink over several enriched and fascinating frameworks developed in last few decades\,\cite{Harrison:2002er, Ma:2004zv, Altarelli:2005yp,Altarelli:2005yx, Babu:2005se, deMedeirosVarzielas:2005qg, Ma:2001dn, Babu:2002dz, Grimus:2008vg, Mohapatra:2006pu, Lam:2008rs, Bazzocchi:2008ej, Feruglio:2007uu,Carr:2007qw, deMedeirosVarzielas:2006fc,Ishimori:2008uc}. Several models based on discrete flavor symmetries which posit a vanishing reactor angle, apparently sound less credible after the reactor angle is proclaimed to be large\,\cite{Tortola:2012te, Forero:2014bxa} and equivalent to the Cabibbo angle ($\theta_{C}$)\,\cite{cabibbo:1978nk}, in quark sector. The Bi-maximal (BM) framework\,\cite{Barger:1998ta, Tanimoto:1998yz, Jezabek:1998du, Nomura:1998gm,Davidson:1998bi,Kang:1998gs,Jarlskog:1998uf,Shafi:1999rm,Fritzsch:1999im,Scott:1999bs, Xing:1999wz,Shafi:1999au,King:2000ce} is one such example that predicts zero reactor angle and maximal solar angle. But both the predictions are now null and void except the prediction of a maximal $\theta_{23}$ which is still consistent within $1\sigma$ error. This indicates either to preclude the BM framework or to harmonize the underlying motivation a little. One such improved follow-up of the BM mechanism is the Bi-Large scenario\,\cite{Boucenna:2012xb,Ding:2012wh, Branco:2014zza, Roy:2012ib,Roy:2014nua}which assumes the solar and the atmospheric angles as equal, non-maximal, and large in general, but leaves a scope to embrace a partial Bi-maximal scenario. Also, in this framework the reactor angle is assumed as large as the Cabibbo angle.     
  
The neutrino mixing, in general, is characterized by six parameters : three mixing angles, $\theta_{12}$\,(solar angle), $\theta_{13}$\,(reactor angle) and $\theta_{23}$\,(atmospheric angle), followed by three phases\,: one Dirac type ($\delta_{CP}$) and two Majorana type CP violating phases ($\kappa $, $\gamma$). The neutrino oscillation experiments witnesses all the observational parameters mentioned above, except the two Majorana phases. All these parameters are contained within,\,$U$,\,the so called Pontecorvo-Maki-Nakagawa-Sakata (PMNS) matrix or lepton mixing matrix which under Standard parametrization, appears as shown,
\begin{equation}
U=R_{23}(\theta_{23})U_{13}(\theta_{13}:\delta)R_{12}(\theta_{12}).P,
\end{equation}
where, $P$ is the diagonal matrix that shelters the two Majorana phases. In \textit{symmetry} basis, where both charged lepton and neutrino mass matrices are considered as non-diagonal, one can identify the lepton mixing matrix, $U=\mathcal{U}_{lL}^{\dagger}.U_{\nu}$\,\cite{King:2013eh, Antusch:2005kw,Romanino:2004ww,Altarelli:2004jb,Plentinger:2005kx, Duarah:2012bd,Roy:2013yka}, where, $\mathcal{U}_{lL}$ is the left handed charged lepton diagonalizing matrix and $U_{\nu}$ is the neutrino mixing matrix. If it is the \textit{flavor} basis, only charged lepton mass matrix is diagonal, and one sees, $U=U_{\nu}$.

As per the BM ansatz, $U_{BM}=R_{23}(\pi/4)R_{12}(\pi/4)$. The specific choice of PMNS matrix, $U=U_{BM}$, highlights the working basis as flavor one, but once redirected to symmetry basis , $U_{BM}$ is open to further amendment from charged lepton sector, and $U=\mathcal{U}_{lL}^{\dagger}U_{BM}$. At the same time, the choice of both $\mathcal{U}_{lL}$ and charged lepton mass matrix turn significant. In this respect, we shall be headed by certain GUT motivated phenomenology which highlights the possible kinship between ``down-quark'' and ``charged-lepton'' mass matrices, and shall also discuss a little on the scenario, the tie-in is slacken. The first scenario leads to definite choices of $\mathcal{U}_{lL}$, whereas the second approach is little conjectural and asks for concrete conceptual rationale.

Here, we emphasize that the strides undertaken in the present approach differs from those where the original chassis is mended either from charged lepton or neutrino sector, rather we adopt the situation where both sectors share a part\citep{Ahn:2011if,Girardi:2015vha}.

The present analysis endeavors to identify a predictive frame work based on BM mixing mechanism, and to probe those prospects that may lead the neutrino mixing towards the unified picture of flavors. 
\input{prdinput0}
\section{BM and BM deviated scenarios}
\subsection{The neutrino mixing matrix}
We shall part the discussion on PMNS matrix deviated from BM mixing in three rostrums: \textbf{Scheme-I}, \textbf{Scheme-II} and \textbf{Scheme-III}. First we choose the flavor basis. Up to the Majorana phases, three versions of the neutrino mixing matrices are presented as in the following,
\begin{itemize}
\item [(i)]\textbf{Scheme-I}
\begin{equation}\label{Ubm1}
U_{\nu}=U_{BM}= R_{23}\left( -\frac{\pi}{4}\right)R_{12}\left( \frac{\pi}{4}\right) ,
\end{equation}
where the sign convention undertaken is in accordance with the ref.\,\cite{Harrison:2002er}. The present texture of $U_{\nu}$ just highlights the original BM proposition.
\item[(ii)] \textbf{Scheme-II}

Multiplying the $U_{BM}$ further with a boost $W_{12}^{T}(\epsilon)$ from right, we obtain,
\begin{eqnarray}
\label{Ubm2}
U_{\nu}&=&U_{BM}.W_{12}^{T}(\epsilon),\\
       &=& R_{23}\left( -\frac{\pi}{4}\right)R_{12}\left( \frac{\pi}{4}-\epsilon\right)
\end{eqnarray}
where,
\begin{equation}
 W_{12}\approx  \left[
\begin{array}{ccc}
 1-\frac{\epsilon ^2}{2} & \epsilon  & 0 \\
 -\epsilon  & 1-\frac{\epsilon ^2}{2} & 0 \\
 0 & 0 & 1 \\
\end{array}
\right], 
\end{equation}
where, $\epsilon$ is an unknown parameter. The above texture emphasizes on the possible modulation of BM mixing from $1$-$2$ sector.
\item[(iii)] \textbf{Scheme-III}

Another speculation which associates a nonzero reactor angle, $\theta_{13}\sim\mathcal{O}(\theta_{C})$ with original BM framework is presented below, 
%\begin{widetext}
\begin{eqnarray}
\label{Ubm3}
U_{\nu}= R_{23}\left( \frac{\pi}{4}\right)U_{13}\left(\alpha\theta_{C}:\delta \right) R_{12}\left( \frac{\pi}{4}\right)
\end{eqnarray} 
%\end{widetext}
Where, $\alpha$ is an unknown $\mathcal{O}(1)$ coefficient. The above design is motivated in the Bi-large mixing frameworks\,\cite{Boucenna:2012xb,Roy:2012ib,Roy:2014nua} and it finds some resemblance with Tri-Bimaximal Cabibbo mixing matrix\,\cite{King:2012vj}. Unlike the original BM mixing scheme (\textbf{Scheme-I}), the present one permits the entry of Dirac-CP violating phase within the neutrino sector. The parameter $\delta$ is also free. On $\alpha$ being zero, \textbf{Scheme-III} coincides with \textbf{Scheme-I}.
\end{itemize}
The detailed textures of $U_{\nu}$'s corresponding to \textbf{Schemes-I,\,II} and \textbf{III} are highlighted in Table\,\ref{table0}.  

\subsection{The neutrino mass matrix}
It is pertinent to trace out the textures of the neutrino mass matrices in \textbf{Schemes-II-III} which are necessary to understand the mechanism of symmetry and the break down of the same as well.
\begin{itemize}
\item[(i)] We know that \textbf{Scheme-I} shows a $S_{3}$ invariant texture  as shown below,
\begin{eqnarray}
\label{mutau}
M_{BM}=\left[
\begin{array}{ccc}
 x & y & y \\
 y & z & x-z \\
 y & x-z & z \\
\end{array}
\right].
\end{eqnarray}
The above texture follows $2$-$3$ interchange symmetry\,[\,also called $\mu$-$\tau$ symmetry\,]. 
\item[(ii)] If it is \textbf{Scheme-II}, the texture in Eq.\,(\ref{mutau}) is muddled but the $2$-$3$ symmetry is fortified. We express its texture,
\begin{eqnarray}
\label{Mbm2}
M_{\mu\tau}'&\approx & M_{BM}+\sqrt{2} y \epsilon \left[
\begin{array}{ccc}
 -2 & 0 & 0 \\
 0 & 1 & 1 \\
 0 & 1 & 1 \\
\end{array}
\right]\nonumber\\
&&-2 y \epsilon^2 \left[
\begin{array}{ccc}
 0 & 1 & 1 \\
 1 & 0 & 0 \\
 1 & 0 & 0 \\
\end{array}
\right]\nonumber\\
&&-\frac{1}{\sqrt{2}} y \epsilon ^3 \left[
\begin{array}{ccc}
 -2 & 0 & 0 \\
 0 & 1 & 1 \\
 0 & 1 & 1 \\
\end{array}
\right].
\end{eqnarray}  
\item[(iii)] But, in \textbf{Scheme-III}, the $\mu$-$\tau$ symmetry is breached. However, this is interesting to note that the neutrino mass matrix as a whole reflects a blend of a $\mu$-$\tau$ and anti-$\mu$-$\tau$ symmetric textures. Let. for simplicity, when $\delta=0$, one can enunciate,
\begin{widetext}
\begin{eqnarray}
\label{Mbm3}
 M_{\nu} &=&
  \left[
\begin{array}{ccc}
 x & y & y \\
 y & z & x-z \\
 y & x-z & z \\
\end{array}
\right] + \sqrt{2}a\lambda \left[
\begin{array}{ccc}
 0 & x-z & z-x \\
 x-z & y & 0 \\
 z-x & 0 & -y \\
\end{array}
\right]+ a^2\lambda^2 \left[
\begin{array}{ccc}
 2(z-x) & -\frac{y}{2} & -\frac{y}{2} \\
 -\frac{y}{2} & x-z & z-x \\
 -\frac{y}{2} & z-x & x-z \\
\end{array}
\right]\nonumber\\
&&+\frac{1}{\sqrt{2}}a^3\lambda^3 \left[
\begin{array}{ccc}
 0 & z-x & x-z \\
 z-x & 0 & 0 \\
 x-z & 0 & 0 \\
\end{array}
\right]+ \frac{a^4\lambda^4}{8}\left[
\begin{array}{ccc}
 2 x & 0 & 0 \\
 0 & 2 z-x & x-2 z \\
 0 & x-2 z & 2 z-x \\
\end{array}
\right]+\mathcal{O}(\lambda^5).
\end{eqnarray}
\end{widetext}
This is interesting to note that textures associated with even powers of $\lambda$ are $\mu$-$\tau$ symmetric and those for odd powers are anti $\mu$-$\tau$ symmetric.
\end{itemize}
We see that \textbf{Schemes-II-III}, although encompass the possible amendments either in terms of $\theta_{12}$ or $\theta_{13}$, yet fail to describe a complete picture. Hence, we need to redefine the Schemes in \textit{symmetry} basis. But before that we investigate the possible forms of $\mathcal{U}_{lL}$s inspired in Grand unified theories\,(GUT). 
\input{prdinput1}

\section{Textures of $\mathcal{U}_{lL}$ from SU(5) GUT}

Though the lepton and quark sectors differ a lot from the mixing point of view, the SO(10) and SU(5) GUTs reflects rational possibilities to link the two sectors upto certain extent. In GUT, a single joint operator can engender the elements of both quark and lepton Yukawa matrices. This signifies a possible link-up between the Yukawa matrices for ``down-type'' quarks ($Y_{d}$) and ``charged'' leptons ($Y_{l}$) in terms of certain ``GUT'' motivated relations. 

For example, the Pati-Salam \,(PS) models posit: $Y_{l}\simeq Y_{d}$\,\cite{Pati:1974yy,Elias:1975kf,Elias:1977bv, Blazek:2003wz,Dent:2007eu}. If, $Y_{e}$ is exactly equal to $Y_{d}$, then one can directly equate $\theta_{12}^{l}$ to $\theta_{12}^{d}\,(\simeq \theta_{C})$.  This at the same time says, $\mathcal{U}_{lL}=U_{d}\simeq V_{CKM}$.

On the other hand, SU(5) models reveal, $Y_{l}\simeq Y_{d}^{T}$\,\cite{kounnas1984grand,ross2003grand}. Following the road-maps of the refs.\,\cite{Antusch:2009gu,Marzocca:2011dh,Antusch:2012fb,Antusch:2013ti,Antusch:2013rxa} we develop certain SU(5) inspired textures of $Y_{l}$ which describes, 
\begin{equation}
\sin\theta_{12}^{l}=\beta\,\lambda.
\end{equation}
Where, $\lambda=\sin\theta_{C}\simeq \theta_{C}$ and $\beta$ encompasses the possibilities: both $\beta \gtrsim 1$, and $\beta \lesssim 1$ [see Table\,\ref{table1}]. We summarize the steps undertaken. According to SU(5) GUT, if
\begin{eqnarray}
\label{SU5}
Y_{d}&=&\begin{bmatrix}
d & b &0\\
a & c & 0\\
0 & 0 & f
\end{bmatrix},
\,\text{then}\,\\
Y_{l}&=&\begin{bmatrix}
\mathcal{A}_{11}\,d & \mathcal{A}_{12}\,b & 0\\
\mathcal{A}_{21}\,a & \mathcal{A}_{22}\,c & 0\\
0  & 0 & \mathcal{A}_{33}\,f 
\end{bmatrix}^{T}.
\end{eqnarray}
The coefficients $\mathcal{A}_{ij}$'s are driven by the common joint operator, from where $Y_{d}$ and $Y_{l}$ emerge. With the dimension $\mathbf{5}$ operator, $\mathcal{A}_{ij}$'s have following choices,
\begin{eqnarray}
\label{coeffs}
\mathcal{A}_{ij} =\left\lbrace \frac{1}{6}, -\frac{1}{2},-\frac{2}{3},1,\pm\frac{3}{2},-3,\frac{9}{2},6,9,-18 \right\rbrace.
\end{eqnarray}
The $\mathcal{U}_{lL}$ is the matrix that diagonalize $Y_{l}.Y_{l}^{\dagger}$, and for above texture, it assumes the form,
\begin{eqnarray}
\label{ulsu5}
\mathcal{U}_{lL}\simeq \begin{bmatrix}
1-\frac{\beta\,\lambda^2}{2} & \beta\,\lambda & 0 \\ 
-\beta\,\lambda & 1-\frac{\beta\,\lambda^2}{2} & 0 \\ 
0 & 0 & 1
\end{bmatrix},
\end{eqnarray}
under the assumption, $\lbrace a,\,b,\,c,\,d \rbrace$ are all real. The parametrization of these quantities and selection of $\mathcal{A}_{ij}$'s must accompany consistent prediction of fermion mass ratios and $|V_{us}|$\,\cite{Antusch:2013rxa,Antusch:2013jca,Charles:2004jd}:
\begin{eqnarray}
\left|\frac{y_{\mu}}{y_{s}}\right| & \approx & 9/2,\,6,\quad \left|\frac{y_{\tau}}{y_{b}}\right|\approx 3/2,\\
\left|\frac{y_{\mu} y_{d}}{y_{s} y_{e}}\right| & \approx & 10.7^{+1.8}_{-0.8},\quad\left|V_{us}\right|  \approx  0.2255. 
\end{eqnarray}

To illustrate, let  
\begin{eqnarray}
Y_{d}&\sim &\begin{bmatrix}
d & b &0\\
a & c & 0\\
0 & 0 & f
\end{bmatrix},\\
Y_{l}&\sim &\begin{bmatrix}
0 & -\frac{2}{3}\,b & 0\\
\frac{9}{2}\,a & 6\,c & 0\\
0 & 0 & -\frac{3}{2}\,f
\end{bmatrix}^{T},
\end{eqnarray}
and with $\lbrace a,b,c \rbrace=\lbrace 0.24,0.244,1 \rbrace$ it predicts, 
\begin{equation}
\sin\theta_{12}^l\approx 0.785\,\lambda,
\end{equation}
and satisfies all the necessary condition. For more details, we refer to Table\,\ref{table1} where five other possibilities are also highlighted. In the present context, we are interested in the GUT motivated relation, $|y_{\mu}/y_{s}|\approx 6$, and the textures of $Y_{l}$'s are designed accordingly. A more rigorous treatment on this issue is available in ref\,\cite{Marzocca:2011dh}; but the present discussion includes those possibilities like $\mathcal{A}_{ij}=1/6,\,-2/3$ which were predicted later in ref.\,\cite{Antusch:2013rxa} and are unfounded in the former.  

In literature we very often encounter the scenarios like: Georgi-Jarlskog\,(GJ) mechanism leads, $\beta\approx 1/3$\,\cite{Georgi:1979df}; also, $\beta$ may assume a value $1$\,\cite{Antusch:2012fb}; several SUSY breaking schemes, like mGSMB and CMSSM assign $\beta$ certain fractional values like $1/6$ and $2/9$ respectively\,\cite{Antusch:2009gu}. Another possibility is found in ref\,\cite{Antusch:2005ca} where certain operators generating fermion masses may lead to,\,$\beta=3/2$. 
 
\input{prdinput2}
\subsection{``CKM-like'' texture}
The above discussion contributes a lot to delineate the texture of $\mathcal{U}_{lL}$. The, $\mathcal{U}_{lL}$, being a $3\times 3$, unitary matrix, requires six phases, $\phi_{ij}$ s in addition to three angular parameters, to parametrize the same. This motivates one to define a  generalized ``\textbf{\textit{CKM-like}}'' texture with, 
\begin{equation}
\theta_{12}^{l}\approxeq \beta\,\lambda,\hspace{0.5 cm}\theta_{23}^{l}\approx A\,\lambda^2,\hspace{0.5 cm} \theta_{13}^{l}\sim \lambda^3,\nonumber
\end{equation} 
as shown below,
\begin{eqnarray}
\label{ul1}
\mathcal{U}_{lL}\approx U_{12}\left( \beta\lambda\,: \phi_{12}\right). U_{13}\left( A\lambda^3:\,\phi_{13}\right) U_{12}\left(A\lambda^2\,: \phi_{23}\right)\nonumber\\
\end{eqnarray}
Here, $\lambda$, $A$, $\rho$ and $\eta$ are the standard Wolfenstein parameters\,\cite{PhysRevLett.51.1945}. Besides the unknown phases $\phi_{12}$ and $\phi_{23}$, there are three extra phases which are quenched on redefining the three right-handed charged lepton fields. The \textbf{CKM-like} $\mathcal{U}_{lL}$ is relevant especially for those scenarios, where the neutrino sector perceives least information of Dirac-type CP violation. 

\subsection{``Close to CKM texture''}
Once we portray $\mathcal{U}_{lL}$ as $V_{CKM}$ or close to $V_{CKM}$, beside the similarity of angles the similarity of the phases become important. The subsequent choices of $U_{eL}$ respect this stand. We put $\phi_{12}=0$ and let $\phi_{23}=0$ or $\pi$.
\begin{eqnarray}
\label{ul2}
\mathcal{U}_{lL}\approx R_{12}\left( \beta\lambda\right). U_{13}\left( A\lambda^3:\,\phi_{13}\right) R_{12}\left(\pm\,A\lambda^2\right).
\end{eqnarray}
On neglecting the small CKM type $1$-$2$ and $2$-$3$ rotational effects, the above texture coincides exactly with the original SU(5) texture of $\mathcal{U}_{lL}$ in Eq.\,(\ref{ulsu5}).  
\subsection{``Exact-CKM texture''}
Additionally, if $\beta=1$, which suggests $Y_{e}=Y_{d}$, we encounter, $\mathcal{U}_{lL}=V_{CKM}$.
\begin{eqnarray}
\label{ul3}
\mathcal{U}_{lL}\approx R_{12}\left( \lambda\right). U_{13}\left( A\lambda^3:\,\phi_{13}\right) R_{12}\left(\pm\,A\lambda^2\right).
\end{eqnarray}
In principle, the contribution of $U_{13}(A\lambda^3\,:\,\phi_{13})$ can be neglected. But its presence may highlight the small CKM like CP contribution towards the lepton mixing matrix in terms of $\phi_{13}$, where,
\begin{equation}
\phi_{13}=-\tan^{-1}\frac{\eta}{\rho}.
\end{equation} 
The Table\,\ref{table2} contains the details of the above textures.

\section{Symmetry basis}
Now we shall redefine \textbf{Schemes-I, II} and \textbf{III} in symmetry basis. For all numerical analysis or comparison, we shall adhere to the ref\,\cite{Forero:2014bxa}. 
\subsection{Scheme-I and CP conservation}
In \textbf{Scheme-I} [see Eq.\,(\ref{Ubm1})] the association of \textbf{\textit{close to CKM}} type $\mathcal{U}_{lL}$ [see Eq.\,(\ref{ul2})] with the existing $U_{\nu}$, brings about,
\begin{eqnarray}
\theta_{12} &\approx & \frac{\pi}{4}-\frac{\beta \lambda }{\sqrt{2}}+\frac{A \beta \lambda ^3}{\sqrt{2}}-\frac{\beta^3 \lambda ^3}{3 \sqrt{2}}-\frac{A \lambda ^3 \rho }{\sqrt{2}},\\
\theta_{13} &\approx & \frac{\beta \lambda }{\sqrt{2}},\\
\theta_{23} &\approx & \frac{\pi }{4}+\lambda ^2 \left(A-\frac{\beta^2}{4}\right),\\
\delta_{CP} &\approx & n\,\pi + \frac{A \eta  \lambda ^2}{\beta}.
\end{eqnarray}
We see that, $\theta_{12}\approx\frac{\pi}{4}-\theta_{13}$. So if  $\theta_{13}$ rises, then  $\theta_{12}$ will go down. But $\theta_{13}$ is not free and is dominated by the model dependent parameter $\beta$. To obtain best results for $\theta_{12}$, $\beta>1$, and that for $\theta_{13}$ requires, $\beta<1$. So in this scheme best possibility is to choose the limiting condition, $\beta=1$ which reveals,
\begin{eqnarray}
\theta_{12}&\approx & 36^{0}\,(2\sigma), \quad \theta_{13}\approx 9.17^0\, (1\sigma),\\
\theta_{23} &\approx & 46.64^0\,(1\sigma),\quad \delta_{CP} \approx \pi.
\end{eqnarray}
As another possibility, we associate ``\textbf{CKM-like}'' $\mathcal{U}_{lL}$ [see Eq.\,(\ref{ul1})] with \textbf{Scheme-I}. With this modified set-up the  oscillation observables appear as in the following, 
\begin{eqnarray}
\theta_{12} &\approx & \frac{\pi}{4}-\frac{\beta \lambda }{\sqrt{2}}\cos\phi_{12}+\frac{A\beta \lambda ^3}{\sqrt{2}}\cos \left(\phi _{12}+\phi _{23}\right)\nonumber\\
&&-\frac{\beta^3 \lambda ^3}{3 \sqrt{2}}\cos^3\phi_{12}-\frac{A \lambda ^3 \rho }{\sqrt{2}},\\
\theta_{13} &\approx & \frac{\beta \lambda }{\sqrt{2}},\\
\theta_{23} &\approx & \frac{\pi }{4}+\lambda ^2 \left(A\cos\phi _{23}-\frac{\beta^2}{4}\right),\\
\delta_{CP} &\approx & n\,\pi -\phi_{12}+\frac{A\,\lambda^2}{\beta}\left( \beta \sin \phi _{23}-\eta  \cos \phi _{12}+\rho  \sin \phi _{12}\right).\nonumber\\
\end{eqnarray}
In contrast to the previous scenario predictions now involve two angular parameters $\lbrace\phi_{12}, \phi_{23}\rbrace $ which are constrained within $0$ and $\pi$. But parametrization of both the unknowns with respect to the observables is difficult. Let us choose, $\beta=1.03452$ [see Table\,\ref{table1}] and apply a condition, $\phi_{12}+\phi_{23}=90^0$, so that $\theta_{12}$ is depleted maximally from $45^{0}$. Let, $\phi_{12}=0$, and one sees that $\theta_{12}$ reaches,
\begin{equation}
\theta_{12}\approx 35.32^0 \,(1\sigma)< \sin^{-1}\left( \frac{1}{\sqrt{3}}\right),
\end{equation} 
The other observables are predicted as in the following,
\begin{eqnarray}
\theta_{13}\approx 9.49^{0}\,(2\sigma),\theta_{23}\approx 44.22^0\,(1\sigma),\,\delta_{CP}\approx 0.99\,\pi.
\end{eqnarray}
The similar treatment if conducted with $\beta=1$, begets,
\begin{eqnarray}
\theta_{12}&\approx & 35.65^{0}\,(2\sigma), \quad \theta_{13}\approx 9.17^0\, (1\sigma),\\
\theta_{23} &\approx & 44.27^0\,(1\sigma),\quad \delta_{CP} \approx 0.99\pi.
\end{eqnarray}

We conclude that \textbf{Scheme-I} depicts only the CP suppressed scenarios and highlights both the possibilities: $\theta_{23}>45^{0}$ and $\theta_{23}<45^0$. At a time, either $\theta_{12}$ or $\theta_{13}$ can be predicted more precisely than the other. The \textbf{Scheme-I} is simple and hardly uses any observational parameters as input. It is also interesting to note that with appropriate choice of $\beta$, the solar angle can be lowered even from TB prediction.

\subsection{Scheme-II and modified QLC relation}
The \textbf{Scheme-II}, $\theta_{12}$ is subjugated mostly form neutrino sector with unknown parameter, $\epsilon$, and an extension of \textbf{Scheme-II} in the light of ``\textbf{CKM-like}'' $\mathcal{U}_{lL}$  begets the following sum rules, 
\begin{eqnarray}
\label{sc11}
\theta_{12} &\approx & \frac{\pi}{4}-\frac{ \beta \lambda }{\sqrt{2}}\cos\phi_{12}+\frac{A\beta^3 \lambda ^3}{\sqrt{2}}\cos \left(\phi _{12}+\phi _{23}\right)-\frac{A \lambda ^3 \rho }{\sqrt{2}}\nonumber\\
& &-\frac{ \beta^3 \lambda ^3}{3 \sqrt{2}}\cos^3\phi_{12}-\epsilon  \left(\beta^2\lambda ^2 \cos ^2\phi _{12}-\beta^2\lambda ^2+1\right),\nonumber\\
\label{sc12}
\theta_{13} &\approx & \frac{\beta\lambda }{\sqrt{2}},\\
\label{sc13}
\theta_{23} &\approx & \frac{\pi }{4}+\lambda ^2 \left(A\cos\phi _{23}-\frac{\beta^2}{4}\right),\\
\label{sc14}
\delta_{CP} &\approx & n\,\pi -\phi_{12}+\frac{A\,\lambda^2}{\beta}\left( \sin \phi _{23}-\eta  \cos \phi _{12}+\rho  \sin \phi _{12}\right).\nonumber\\
\end{eqnarray}
Unlike \textbf{Scheme-I}, the prediction of $\theta_{12}$ depends a little on $\theta_{13}$.
Here, we find three free parameters: $\lbrace\epsilon,\phi_{12},\phi_{23} \rbrace $. To illustrate, let $\beta=1$. For simplicity, we assume the maximal deviation of $\theta_{23}$ from $45^0$ which implies $\phi_{23}=(2n+1)\pi/2$, So, one sees either, 
\begin{equation}
\theta_{23}\approx 41.9^0\,(1\,\sigma)\quad \text{or} \quad \theta_{23}\approx 46.64^{0}\,(1\,\sigma).
\end{equation} 
The related sum rule is approximated as in the following,
\begin{equation}
\theta_{23}\pm A\lambda^2\approx \frac{\pi}{4}-\theta_{13}^2.
\end{equation}
Let us visualize the situation when CP violation is maximum. It reveals, $\phi_{12}\approx 0.5 \,\pi$, and on choosing of $\epsilon\approx \theta_{C}$, one sees,
\begin{equation}
\theta_{12}\approx 32.62^0\,(3\sigma)< \sin^{-1}\left( \frac{1}{\sqrt{3}}\right),
\end{equation}   
followed by a sum rule upto $\mathcal{O}(\lambda^{2})$,
\begin{eqnarray}
\label{qlc1}
\theta_{12}+\theta_{C}\approx\frac{\pi}{4},
\end{eqnarray}
which is the original Quark lepton complementarity (\textbf{QLC}) relation\,\cite{Raidal:2004iw,Minakata:2004xt,Ferrandis:2004vp,Zhang:2012xu,Antusch:2005ca}. But in order to acquire a precise $\theta_{12}$, we deviate a little from the condition of maximal CP violation. On choosing $\phi_{12}\approx 0.567\pi$ we obtain,
\begin{eqnarray}
\theta_{12}\approx 34.62^{0}\,(\text{central value}),\quad\delta_{CP}\approx 1.43\,\pi \,(1\sigma),
\end{eqnarray}  
which one can relate with a new version of \textbf{QLC} relation\,[\,obtained from Eqs.\,(\ref{sc11}),(\ref{sc12}) and (\ref{sc14})] as highlighted below,
\begin{eqnarray}
\label{qlc}
\theta_{12}+\theta_{c}&\approx & \frac{\pi}{4}-\theta_{13}\cos(n\pi-\delta_{CP}).
\end{eqnarray}
This is to be noted that, Eq.\,(\ref{qlc1}) is obtainable only when $\delta_{CP}=(2n+1)\pi/2$. The reactor angle depends only on $\beta$ and for this special case, when $\beta=1$, it is predicted as,
\begin{equation}
\theta_{13}\approx 9.17^{0}\,(1\sigma).
\end{equation}
With different choice of $\beta$, further lowering of the same is possible. The details of the present scheme are sorted in Table\,\ref{table3}. 
\input{prdinput3}

Let us summarize the possibilities with \textbf{Scheme-II}. 
\begin{itemize}
\item[(i)]In contrast to \textbf{Scheme-I}, the prediction of the results are more precise and are consistent within $1\sigma$ range. The strife between $\theta_{12}$ and $\theta_{13}$ is tamed. 
\item[(ii)]The parametrization concerns three free parameters. The observable, $\theta_{12}$ is chosen as input, and $\phi_{23}$ is fixed either at $0$ or $\pi$. 
\item[(iii)]The interesting feature of \textbf{Scheme-II} is that it hoists the QLC relation in revised form and the original form is reinstated if CP violation is maximum. In view of this, the choice of the free parameter, $\epsilon$ as $\theta_{C}$ is relevant. 
\item[(iv)] The \textbf{Scheme-II} does not advocate for CP suppressed cases, but in order to obtain precise $\theta_{12}$, it depicts a CP violation shifted a little from maximality. 
\end{itemize} 
\input{prdinput4}
\input{prdinput5}
\subsection{Scheme-III and large $\mathbf{\theta_{13}^{\nu}\,\sim\mathcal{O}(\theta_{c})}$}
For, \textbf{Scheme II}, the prediction of observable CP violation is solely dependent on the charged lepton sector. But if it is \textbf{Scheme-III}, this dependency is subdued. In the present scheme, we shall concentrate mostly on the parametrization, where the neutrino sector leads the CP violation.

We first concentrate to the generalized extension of \textbf{Scheme-III} where both charged and neutrino sector contribute toward observable CP violation. This framework embraces a ``\textbf{CKM-like}'' $\mathcal{U}_{lL}$ [see Eq.\,(\ref{ul1})], and one sees that the concerned observational parameters are headed by four unknown parameters: $(\alpha,\,\delta)$ from neutrino sector and $(\phi_{12}, \phi_{23})$ from charged lepton sector. We see the sum rules appear as shown in the following,
\begin{widetext}
\begin{eqnarray}
\theta_{12} &\approx & \frac{\pi }{4}-\frac{\beta \lambda  \cos \phi _{12}}{\sqrt{2}}+\lambda^3 \left(\frac{1}{2} \alpha \beta^2 \cos \delta-a \beta^2 \cos\phi_{12} \cos (\phi_{12}-\delta ) +\frac{A b \cos (\phi_{12}+\phi_{23})}{\sqrt{2}}-\frac{A \rho }{\sqrt{2}}-\frac{\beta^3 \cos ^3\phi_{12}}{3 \sqrt{2}}-\frac{\alpha^2 \beta \cos \phi_{12}}{2 \sqrt{2}}\right),\nonumber\\
\\
\theta_{13} &\approx & \theta_{c} \sqrt{\alpha^2+\sqrt{2} \alpha \beta \cos \left(\delta -\phi _{12}\right)+\frac{\beta^2}{2}}+\frac{1}{6} \lambda ^3 \left(\alpha^2+\sqrt{2} \alpha \beta \cos \left(\delta -\phi _{12}\right)+\frac{\beta^2}{2}\right){}^{3/2},\\
\theta_{23} &\approx & \frac{\pi }{4}+\lambda ^2 \left(A \cos \phi _{23}-\frac{\alpha \beta \cos \left(\delta -\phi _{12}\right)}{\sqrt{2}}-\frac{\beta^2}{4}\right)
\end{eqnarray}
\begin{eqnarray}
\delta_{CP} &\approx & n\pi-\tan ^{-1}\left(\frac{2 \alpha \sin \delta+\sqrt{2} \beta \sin \phi _{12}}{2 \alpha \cos \delta +\sqrt{2} \beta \cos \phi _{12}}\right) -\frac{\lambda ^2}{2 \left(2 \alpha^2+2 \sqrt{2} \alpha \beta \cos \left(\phi _{12}-\delta \right)+\beta^2\right)}\times \lbrace 
-\sqrt{2} \alpha^3 \beta \sin \left(\phi _{12}-\delta \right)\nonumber\\
&& +\sqrt{2} \alpha (2 A (\beta \sin (\phi _{12}+\phi _{23} -\delta)
-\eta  \cos\delta+\rho \sin\delta))+ \beta^3\sqrt{2} \alpha  \sin \left(\phi _{12}-\delta \right)-2 A \beta \eta  \cos \phi _{12}\nonumber\\
&&+2 A \beta \left(\beta \sin \phi _{23}+\rho  \sin \phi _{12}\right)\rbrace.
\end{eqnarray}
\end{widetext}
Let, $\lbrace\phi_{12},\phi_{23},\delta\rbrace\neq 0$. For the present parametrization, $\beta=3/2$ \,\cite{Antusch:2005ca} is found the most suitable one. Here the number of free parameters are equal to that of observational ones. On assigning the angular parameters to their central values $\theta_{12}=34.63^{0}$, $\theta_{13}=8.87^{0}$,\,$\theta_{23}=48.85^{0}$ and $\delta_{CP}\approx 1.32\,\pi$ one obtains, $\phi_{12}\approx 0.23\pi$, $\phi_{23}\approx0.032\,\pi$, $\delta=-0.96\,\pi$ and $a\approx1.23$. One can see that the parametrization reflects an inherent CP suppressed scenario\,($\delta\sim-\pi$) and substantiates a large $1$-$3$ mixing ($\theta_{13}^{\nu}\sim a\,\lambda$),
\begin{equation}
\theta_{13}^{\nu}\approx 15.6^{0}\,>\theta_{C},
\end{equation}
within the neutrino sector. The present parametrization involves as many free parameters as the observable ones and is less predictive.

Next we shall attend the parametrization of \textbf{Scheme-III} with following possibilities,
\begin{itemize}
\item[(i)]$\phi_{12}=0$, that is charged lepton sector contributes least towards observable CP violation and we expect $\mathcal{U}_{lL}$ to assume a ``\textbf{Close to CKM}'' texture. This involves only two free parameters $\alpha$ and $\delta$ and,
\item[(ii)]in addition, as we expect $\theta_{13}^{\nu}\sim\mathcal{O}(\theta_{C})$ similar to that for $\theta_{12}^{l}$, one can further make the parametrization more predictive with a rational \textit{ansatz}, $\theta_{13}^{\nu}=\theta_{12}^{l}\sim\mathcal{O}(\theta_{C})$ which implies  $\alpha=\beta$. This indicates the involvement of single free parameter $\delta$.
\end{itemize} 
\input{prdinput6}
\input{prdinput7}
To address the first possibility, we consider the observable parameters $\theta_{13}$ and $\delta_{CP}$ as input parameters and $\theta_{13}^{\nu}$ and treat the internal CP phase $\delta$, $\theta_{12}$ and $\theta_{23}$ as the predictions. To illustrate, let $\sin^2\theta_{13}=0.023$ and $\delta_{CP}=1.34\,\pi$ (central values). Say, $\beta=1.07894$. Adopting either of the possibilities, $\phi_{23}=0$, or $\pi$ one sees,
\begin{eqnarray}
\theta_{13}^{\nu}&=&-9.18^0\,(-9.94^0),\nonumber\\
\delta &=& 0.0925\pi\,\,(0.09571\pi),\nonumber\\
\theta_{12} &=& 35.33^{0}\,(34.51^0)\,[1\sigma],\nonumber\\
\theta_{23} &=& 47.55^0\,(42.71^0)\,[1\sigma].
\end{eqnarray}
Here, we see that $|\theta_{13}^{\nu}|\lesssim \theta_{C}$. But, with the same environment, another possibility $|\theta_{13}^{\nu}|\gtrsim \theta_{C}$ along with precise prediction of other observable parameters are also obtainable as shown in the following,
\begin{eqnarray}
\theta_{13}^{\nu}&=&-16.05^0\,(-15.67^0),\nonumber\\
\delta &=& -0.0508\pi\,\,(-0.0545\pi),\nonumber\\
\theta_{12} &=& 35.33^{0}\,(34.51^0)\,[1\sigma],\nonumber\\
\theta_{23} &=& 48.96^0\,(44.11^0)\,[1\sigma].
\end{eqnarray}
One can show that if the model dependent parameter $\beta< 1$, one hardly obtains $|\theta_{13}^{\nu}|>\theta_{C}$. For detailed analysis, we refer to Tables\, \ref{table4} and \ref{table5}.

The same treatment when applied to another possibility, $\delta_{CP}=1.48\pi$ \,(another central value), results in,
\begin{eqnarray}
\theta_{13}^{\nu}&=&-13.01^0\,(-12.40^0),\nonumber\\
\delta &=& 0.074\pi\,\,(0.078\pi),\nonumber\\
\theta_{12} &=& 35.33^{0}\,(34.51^0)\,[1\sigma],\nonumber\\
\theta_{23} &=& 48.18^0\,(43.31^0)\,[1\sigma].
\end{eqnarray}

This is interesting to note that, a condition $|\theta_{13}^{\nu}|\approxeq \theta_{C}$ is reached. In contrast to the previous situation, when input parameter $\delta_{CP}=1.34\pi$, present scenario highlights either of the two possibilities: $|\theta_{13}^{\nu}|\geq \theta_{C}$, or $|\theta_{13}^{\nu}|\leq \theta_{C}$, and never two at a time. The details of the parametrization are highlighted in Table\,\ref{table6}.  
 
Let us concentrate to the second stand, which encompasses the provision, $\alpha=\beta$. This parametrization is the most predictive  in the sense, it uses only one variable, $\delta$ and in order to parametrize it, we fix the observable parameter $\sin^2\theta_{13}=0.023$ as an input. Say, if $\beta=1.07894$, one sees for $\phi_{23}=0,\,\pi$, 
\begin{eqnarray}
\theta_{13}^{\nu}&=& 13.94^0\nonumber\\
\delta &=& 0.25 \pi,\,\nonumber\\
\theta_{12} &=& 35.33^{0}\,(34.51^0)\,[1\sigma],\nonumber\\
\theta_{23} &=& 48.40^0\,(43.66^0)\,[1\sigma],\nonumber\\
\delta_{CP} &=& 1.47\,\pi,\,(1.44\,\pi)\,[1\sigma].
\end{eqnarray}
The other possibilities are accumulated in Table\,\ref{table7}. The present parametrization gives better results for $\beta>1$.  

The \textbf{Scheme-III} has characteristic features which we highlight as in the following . 
\begin{itemize}
\item[(i)] The present scheme vindicates the assumption of $\theta_{12}=\theta_{23}=45^{0}$\,(maximal mixing) as first approximation.
\item[(ii)]We emphasize that $\theta_{13}^{\nu}$ is the output of the parametrization. Interestingly we see that the inherent $1$-$3$ angle within neutrino sector can be, larger $\theta_{13}^{\nu}\sim \pi/10,\pi/20$. In addition to that one can see, $\theta_{13}^{\nu}=\theta_{C}$ and in certain occasion, $\theta_{13}^{\nu}=\theta_{12}^{l}$ also. These two features sound relevant in the context of unified theory of flavors. 
\item[(iii)] In fact, the observable CP violation in the lepton sector may share the contribution both from charged lepton and neutrino sectors in terms of $\phi_{12}$ and $\delta$ respectively, as evident from the approximated generalized expression,
\begin{equation}
\delta_{CP}\approx 2\pi-\tan ^{-1}\left(\frac{2 \alpha \sin \delta+\sqrt{2} \beta \sin \phi _{12}}{2 \alpha \cos \delta +\sqrt{2} \beta \cos \phi _{12}}\right)
\end{equation}
But, once we choose $\mathcal{U}_{lL}$'s with ``\textbf{close to CKM}'' texture and adhere to the $\beta$'s described in Table\,\ref{table1}, we are more close to the original description of $\mathcal{U}_{lL}$'s\,[see Eq.\,(\ref{ulsu5})] motivated in SU(5) GUT. The description negates the presence of $\phi_{12}$. Hence, in this respect, the internal CP phase $\delta$, from neutrino sector plays a promising role. In the present parametrization as $\theta_{13}^{\nu}\neq 0$, this feature is more prominent and seems logical in contrast to those model independent possibilities discussed in refs.\,\cite{Roy:2014nua,Girardi:2015vha}. 

\item[(iv)] The present parametrization is predictive. It uses only one ($\theta_{13}$) or two ($\theta_{13}$ and $\delta_{CP}$) observational parameters as input and predicts the rest and also the two unphysical parameters,\,$\lbrace\theta_{13}^{\nu},\delta\rbrace$.
\end{itemize}

\section{When $\mathbf{Y_{l} \nsim Y_{d}}$}

The discussion so far focuses on the possible patterns of symmetry basis believing $Y_{l}$ and $Y_{d}$ are originated from a single joint operator, whereas another possibility that reinforces a different origin of $Y_{l}$ and $Y_{d}$ is also relevant\cite{Meloni:2010cj}. We add a small extension in this line. We assume that the neutrino sector follows BM mixing and there is no modulation, and the charge lepton sector alone is responsible for all the observable deviation. With this motivation we put forward the following texture zero Yukawa matrix ($Y_{l}$) upto $\mathcal{O}(\lambda^7)$ as in the following,
\begin{equation}
\label{yl1}
Y_{l}\simeq\left[
\begin{array}{ccc}
 2 \lambda ^6 & \frac{\lambda ^3}{\sqrt{2}}\left(1-i\,\frac{1}{\sqrt{2}}\right)  & -\frac{\lambda }{\sqrt{2}}(1+ i\,2\lambda)  \\
0 & \lambda ^2 & \frac{\lambda ^2}{3}\left(1+i\right) \\
0 & -\lambda ^3\left(1+i\,\frac{\lambda }{3}\right) & 1 \\
\end{array}
\right].
\end{equation}
This $Y_{l}$ can be diagonalized with a left-handed diagonalizing matrix, $\mathcal{U}_{lL}$ of which the information are supplied as shown below,
\begin{equation}
|\mathcal{U}_{lL}| \approx  \left[
\begin{array}{ccc}
 0.969 & 0.176& 0.172\\
 0.175 & 0.984& 0.023 \\
 0.173 & 0.013 & 0.984 \\
\end{array}
\right],
\end{equation}
\begin{equation}
{Arg}[\mathcal{U}_{lL}]\approx  \left[
\begin{array}{ccc}
 0.131 & 0.383 & -0.865 \\
 -0.608 & 0.640 & 0.255 \\
 0 & 0 & 0 \\
\end{array}
\right]\pi.
\end{equation}
The right-handed diagonalizing matrix of the above $Y_{l}$ is, $V_{lR}\approx diag\lbrace i,1,1 \rbrace$. Also, $|y_{e}/y_{\mu}|$ is predicted as $0.00494$. The PMNS matrix constructed, $U=\mathcal{U}_{lL}^{\dagger}.R_{23}^{\nu}(\pi/4)R_{12}^{\nu}(\pi/4)$ in this background begets,
\begin{eqnarray}
\theta_{13} &=& 8.17^{0}\,[1\sigma],\\
\theta_{12} &=& 33.52^{0}\,[2\sigma],\\
\theta_{23} &=& 44.35^{0}\,[1\sigma],\\
\delta_{CP} &=& 1.69\,\pi\,[1\sigma].
\end{eqnarray} 
But more important is to trace out the frame work where the texture in Eq.\,(\ref{yl1}) may emerge. Interestingly, we see this texture is encouraged in refs\,\cite{Ludl:2014axa,Ferreira:2014vna}.

\section{Discussion and Summary}

All the $\mathcal{U}_{lL}$'s and the related $Y_{l}$'s discussed in the present article\,[except Eq.\,(\ref{yl1})] are motivated in SU(5) GUT. Similar to the charged lepton sector, it would have been a good exercise to work out the first principle supporting the model independent textures of both $U_{\nu}$ and $M_{\nu}$  highlighted in \textbf{Schemes} \textbf{II} and \textbf{III}. But this is beyond the scope of present article. We wish to discuss the possible link ups that may help the model builders to think in this line, in short.

In the neutrino mass matrix under \textbf{Scheme-II}\,[see Eq.\,(\ref{Mbm2})], the parameter $\epsilon$ is responsible for deviating $M_{\nu}$ from BM mixing scenario within $\mu$-$\tau$ symmetric regime. This phenomenon is somehow akin to flavor twisting effect which is motivated in the extra-dimension inspired frameworks\cite{Haba:2006dz}.

Also, the parameter $\epsilon$ in \textbf{Scheme-II}, leaves a scope to achieve the original QLC relation with little modification [see Table\,\ref{table1} and Eq.\,(\ref{qlc})] by tuning the former to $\theta_{C}$. Perhaps, this  is not just a mere numerical coincidence and one finds the related discussion in the ``Cabibbo Haze'' based theories\,\cite{Datta:2005ci,Kile:2013gla}.    

In \textbf{Scheme-III}, the neutrino mass matrix, $M_{\nu}$ in Eq.\,(\ref{Mbm3}) is approximated as shown below,
\begin{eqnarray}
&& M_{BM}+c_{1}\,\lambda \underbrace{\begin{bmatrix}
0 & -1 & 1 \\ 
-1 & 0 & 0 \\ 
1  & 0 & 0
\end{bmatrix}}+c_{2}\,\lambda \underbrace{\begin{bmatrix}
0 & 0 & 0 \\ 
0 & 1 & 0 \\ 
0  & 0 & -1
\end{bmatrix}}\\
&& \hspace{2.67 cm}\delta m^{type-I}\hspace{1.2 cm}  \delta m^{type-II}\nonumber
\end{eqnarray}
where, $\delta m^{type-I}$ and $\delta m^{type-II}$ resemble the first order perturbation to $M_{BM}$ and possibly, the Type-I and Type-II see-saw mechanisms in the $S_{4}$ symmetric background, may generate these deviation matrices in their respective order\,\cite{PhysRevD.84.053002}. The \textbf{Scheme-III} describes one possibility which in addition to $\mu$-$\tau$ symmetry breaking, requires charged lepton correction also. We see that these methodology is motivated in grand unified theories \cite{Cooper:2012wf, Hagedorn:2012ut}.

In \textbf{Scheme-III}, the situation which highlights $\theta_{13}^{\nu}\sim 18^{0}$  are motivated in refs\,\cite{Bazzocchi:2011ax,Rodejohann:2014xoa}.

Also in \textbf{Scheme-III}, scenarios: $\sin\theta_{13}^{\nu}\simeq  \lambda$, $\sin\theta_{12}^{\nu}= \sin\theta_{23}^{\nu}\simeq 3.13\,\lambda\, (=1/\sqrt{2})$ are inspired in ref\,\cite{Ding:2012wh}. Perhaps the former pattern is  derivable in the Bi-large based frameworks based on $U(1)\times Z_{m}\times Z_{n}$ symmetry, with $m$ and $n$ having different parities.       

The present model independent analysis aspires to refine the BM based framework and tries to relate the same to the unified theory of flavors. The refs.\,\cite{Altarelli:2009gn, Li:2014eia,Petcov:2014laa,Duarah:2015bja} discuss the possibilities to amend the BM framework following other alternatives. The present work finds some similarity with ref.\,\cite{Girardi:2015vha} but the motivations in either cases differ. The latter concerns $\theta_{13}^{\nu}$ as input and assigns preferred values to it and the charged lepton diagonalizing matrices considered therein are arbitrary. But in contrast, the present work considers $\theta_{13}^{\nu}$ as a prediction of a certain parametrization\,[\textbf{Scheme-III}] and encounter several interesting possibilities like $\theta_{13}^{\nu}=\theta_{C}$ and even $\theta_{13}^{\nu}=\theta_{12}^{l}\sim\mathcal{O}(\theta_{C})$ (We hope these relations are important in the context of GUT) in addition to those, $\theta_{13}^{\nu}\sim \pi/10$, $\theta_{13}^{\nu}=\pi/20$ etc. Also, the charged lepton corrections adopted in the present analysis are not arbitrary and inspired in SU(5) GUT.  Also, the present work uses one or two observational parameters as input and sounds more predictive.

To summarize, we have highlighted the new possibilities of $\mathcal{U}_{lL}$'s motivated in SU(5) GUT and have tried to reinstate the BM mixing scheme in terms of modulation, either in $1$-$2$ rotation or $1$-$3$ rotation in the light of charged lepton correction. The parametrization is predictive and hoists a revised QLC relation of which the original one appears as a special case. This scenario however supports a little deviation from maximal CP violation. In addition it spotlights the BM scenarios with $1$-$3$ angle as large as Cabibbo angle, lesser and even larger than the same and also accents the scenarios like $\theta_{13}^{\nu}=\theta_{12}^{l}$. In conclusion, one may infer that the BM mixing which is less attractive in the light of present experimental data, sounds tenable as a first approximation, if the original motivation is tuned a little.

One of the authors, \textbf{SR} wishes to thank Constantin Sluka from University of Basel, Switzerland for the useful discussion.

\bibliography{bmbib}

\end{document}

%% file: prdinput0.tex
\begin{table}
\begin{center}
\setlength{\tabcolsep}{0.5 em}
\begin{tabular}{l|l}
\hline
\hline
\textbf{} & \textbf{Flavor basis} \\
\hline
\hline
\textbf{I} &  
$ \left[
\begin{array}{rrr}
 \frac{1}{\sqrt{2}} & \frac{1}{\sqrt{2}} & 0 \\
 -\frac{1}{2} & \frac{1}{2} & -\frac{1}{\sqrt{2}} \\
 -\frac{1}{2} & \frac{1}{2} & \frac{1}{\sqrt{2}} \\
\end{array}
\right] $
\\
\hline
\textbf{II} &  $\left[
\begin{array}{rrr}
 \frac{1}{\sqrt{2}} & \frac{1}{\sqrt{2}} & 0 \\
 -\frac{1}{2} & \frac{1}{2} & -\frac{1}{\sqrt{2}} \\
 -\frac{1}{2} & \frac{1}{2} & \frac{1}{\sqrt{2}} \\
\end{array}
\right].\left[
\begin{array}{ccc}
 1-\frac{\epsilon ^2}{2} & \epsilon  & 0 \\
 -\epsilon  & 1-\frac{\epsilon ^2}{2} & 0 \\
 0 & 0 & 1 \\
\end{array}
\right]^T  $
\\
\hline
\textbf{III} &  
$ \left[
\begin{array}{crr}
 \frac{1}{\sqrt{2}}\left(1-\frac{\alpha ^2 \lambda ^2}{2}\right) & \frac{1}{\sqrt{2}}\left(1-\frac{\alpha^2 \lambda ^2}{2}\right) & \alpha \lambda e^{-i \delta }   \\
   -\frac{1}{2}\left(1 - \alpha \lambda e^{i \delta }\right) & \frac{1}{2}\left(1+  \alpha \lambda e^{i \delta }\right)  & -\frac{1}{\sqrt{2}}\left(1-\frac{\alpha^2 \lambda ^2}{2}\right) \\
 -\frac{1}{2}\left(1+ \alpha\lambda e^{i \delta }\right)   & \frac{1}{2}\left(1- \alpha \lambda  e^{i \delta }\right)  & \frac{1}{\sqrt{2}}\left(1-\frac{\alpha^2 \lambda ^2}{2}\right) \\
\end{array}
\right] $
\\
\hline
\end{tabular}
\caption{\label{table0}\footnotesize The description of \textbf{Scheme-I,\,II,\,III} in Flavor basis: $U=U_{\nu}$. The \textbf{Scheme -I} depicts the original BM mixing. The \textbf{Schemes-II,} and \textbf{III} describes $U_{\nu}$ deviated from $U_{BM}$ in terms of $\theta_{12}$ and $\theta_{13}$ respectively. }
\end{center}
\end{table}

%% file: prdinput1.tex
\begin{table*}
\begin{center}
\setlength{\tabcolsep}{1.25 em}
\begin{tabular}{llc|c|ccccc}
\hline 
\hline
  &$(Y_{l})_{12}$ & $\lbrace\,d,\,a,\,b\rbrace$ & $\beta$ & $\left|\frac{y_{e}}{y_{\mu}}\right|$ &  $\left|\frac{y_{s}}{y_{d}}\right|$& $|V_{us}|$ & $\left|\frac{y_{\mu}}{y_{s}}\frac{y_{d}}{y_{e}}\right|$  \\ 
\hline 
\hline
(a)& $\left[
\begin{array}{cc}
 9\,d & \frac{3}{2}\,b \\
 3\,a  & 6\,c \\
\end{array}
\right]^{T}$ & $\lbrace 0.0016,\,0.24,\,0.244\rbrace$ & $0.527075$ & $ 0.004832$ & $19.56$& $0.2257$ & $10.57$ \\ 
\hline 
(b)&  $\left[
\begin{array}{cc}
 0 & -\frac{2}{3}\,b\\
 \frac{9}{2}\,a & 6\,c \\
\end{array}
\right]^{T}$ & $\lbrace 0,\,0.24,\,0.244\rbrace$ & $0.785059$ & $0.004723$ & $19.56$ & $0.2257$ & $10.82$ \\ 
\hline 
(c)& $\left[
\begin{array}{cc}
 \frac{1}{6}\,d & -\frac{1}{2}\,b \\
 6\,a & 6\,c \\
\end{array}
\right]^{T}$ & $\lbrace -0.003,\,0.22,\,0.243\rbrace$ & $0.952475$ & $0.004168$ & $19.56$ & $0.2258$ & $12.26$ \\ 
\hline 
(d)& $\left[
\begin{array}{cc}
 -\frac{2 }{3}\,d & -\frac{1}{2}\,b \\
 6\,a & 6\,c \\
\end{array}
\right]^{T}$ & $\lbrace 0.001,\,0.24,\,0.244 \rbrace$ & $1.03452$ & $0.004507$ & $19.35$ &$0.2256$& $11.46$\\ 
\hline 
(e)& $\begin{bmatrix}
1\,d & -\frac{1}{2}\,b\\
6\,a & 6\,c 
\end{bmatrix}^{T}$ & $\lbrace-0.0002,\,0.251,\,-0.245 \rbrace$ & $1.07894$ & $0.004850$ & $18.26$ & $0.2253$ & $11.28$ \\ 
\hline 
(f)& $\left[
\begin{array}{cc}
 \frac{3}{2}\,d & -\frac{1}{2}\,b \\
 9\, a & 6\,c \\
\end{array}
\right]^{T}$ & $\lbrace -0.005,\,0.20,\,0.241\rbrace$ & $1.27392$ & $0.004379$ & $20.59$ & $0.2254$ &$ 11.09$\\
\hline
\end{tabular}
\caption{\label{table1}\footnotesize Different possibilities for $(Y_{e})_{12}$ are illustrated based on SU(5) GUT models. The coefficients appearing in all the matrices allowed by dimension \textbf{5} operator. The above textures respect the GUT motivated relation: $y_{\mu}:y_{s}\approx 6$. The fermion mass ratios are the important parameters in appraising the validity of the above textures. The important parameter $(y_{\mu}/y_{s})(y_{d}/y_{e})$ must lie within $10.7^{+1.8}_{-0.8}$. The above textures highlight different possibilities to parametrize the $1$-$2$ rotation of $U_{lL}$. One can see that all the possibilities including $\beta\gtrsim 1$ or $\beta\lesssim 1$ are allowed, where $\beta=\sin\theta_{12}^{l}/\sin\theta_{C}$. In the above textures, the input parameter $c$ is chosen as unity.} 
\end{center}
\end{table*}

%% file: prdinput2.tex
\begin{table*}
\begin{center}
\setlength{\tabcolsep}{0.4 em}
\begin{tabular}{l|l|l}
\hline
\hline
\textbf{$U_{lL}$} &$\beta$ & Texture\\
\hline
\hline
\textbf{CKM like} & $\beta \neq 1$ & 
$ \left[
\begin{array}{ccc}
 1-\frac{\beta^2 \lambda ^2}{2} & \beta  \lambda e^{-i \phi _{12}}  & A\lambda ^3 (\rho -i\eta ) \\
 -\beta \lambda e^{i \phi _{12}}   & 1-\frac{\beta^2 \lambda ^2}{2} & A \lambda ^2 e^{-i \phi _{23}}  \\
A \lambda ^3 \left(\beta e^{i( \phi _{12}+\phi _{23})}-\rho -i \eta \right) & -A \lambda ^2 e^{i \phi _{23}}  & 1 \\
\end{array}
\right] $
\\
\hline
\textbf{Close to CKM} & $\beta\neq1$ & $ \left[
\begin{array}{ccc}
 1-\frac{\beta^2 \lambda ^2}{2} & \beta  \lambda   & A\lambda ^3 (\rho -i\eta ) \\
 -\beta \lambda    & 1-\frac{\beta^2 \lambda ^2}{2} & A \lambda ^2   \\
A \lambda ^3 \left(\beta -\rho -i \eta \right) & -A \lambda ^2  & 1 \\
\end{array}
\right] $,\,$ \left[
\begin{array}{ccc}
 1-\frac{\beta^2 \lambda ^2}{2} & \beta  \lambda   & A\lambda ^3 (\rho -i\eta ) \\
 -\beta \lambda   & 1-\frac{\beta^2 \lambda ^2}{2} & -A \lambda ^2   \\
-A \lambda ^3 \left(\beta +\rho +i \eta \right) & A \lambda ^2   & 1 \\
\end{array}
\right] $
\\
\hline
\textbf{Exact CKM texture} & $\beta=1$ & 
$\left[
\begin{array}{ccc}
 1-\frac{\lambda ^2}{2} & \lambda   & A\lambda ^3 (\rho -i\eta ) \\
 -\lambda    & 1-\frac{\lambda ^2}{2} & A \lambda ^2   \\
A \lambda ^3 \left(1 -\rho -i \eta \right) & -A \lambda ^2  & 1 \\
\end{array}
\right] $,\,$ \left[
\begin{array}{ccc}
 1-\frac{\lambda ^2}{2} & \lambda   & A\lambda ^3 (\rho -i\eta ) \\
 -\lambda   & 1-\frac{\lambda ^2}{2} & -A \lambda ^2   \\
-A \lambda ^3 \left(1 +\rho +i \eta \right) & A \lambda ^2   & 1 \\
\end{array}
\right] $
\\
\hline
\end{tabular}
\caption{\label{table2}\footnotesize The different choices of $U_{lL}$'s with \textbf{CKM-like}, \textbf{Close to CKM} and \textbf{Exact CKM} textures are depicted. In second and the third textures, both possibilities of $\phi_{23}=0$ and $\phi_{23}=\pi$ are considered.  }
\end{center}
\end{table*}

%% file: prdinput3.tex
\begin{table}
\begin{center}
\setlength{\tabcolsep}{0.4 em}
\begin{tabular}{l|cccc}
\hline
\hline
 $\beta$  & $\frac{\phi _{12}}{\pi }$ & $\theta _{13}({}^0)$ & $\theta _{23}({}^0)$ & $\frac{\delta _{\text{CP}}}{\pi }$ \\
 \hline
\hline
 0.52707 & 0.6642 & 4.82\,$(-)$ & $42.43 -47.17$\,$(1\sigma)$ & 1.33\,$(1\sigma)$ \\
 \hline
 0.78505 & 0.5975 & 7.19\,$(-)$ & $42.18 -46.92$\,$(1\sigma)$ & 1.39\,$(1\sigma)$ \\
 \hline
 0.95247 & 0.5730 & 8.73\,$(1\sigma)$ & $41.95-46.71$\,$(1\sigma)$ & 1.42\,$(1\sigma)$ \\
 \hline
 1 & 0.5674 & 9.17\,$(1\sigma)$ & $41.90-46.64$\,$(1\sigma)$ & 1.43\,$(1\sigma)$ \\
 \hline
 1.03452 & 0.5636 & 9.49 \,$(2\sigma)$& $41.85-46.59$ \,$(1\sigma)$ & 1.43 \,$(1\sigma)$\\
 \hline
 1.07894 & 0.5589& 9.90 \,$(3\sigma)$ & $41.78-46.52$\,$(1\sigma)$ & 1.44\,$(1\sigma)$ \\
 \hline
 1.27392 & 0.5417 & 11.71\,$(-)$& $41.44-46.19$\,$(1\sigma)$ & 1.46\,$(1\sigma)$ \\
 \hline
\end{tabular}
\caption{\label{table3} \footnotesize The predictions of \textbf{Scheme-II} (with $\epsilon=\theta_{C}$) for the observable parameters $\theta_{13}$, $\theta_{23}$ and $\delta_{CP}$ are highlighted. Here $\theta_{12}$ is taken as input parameter: $\sin^2\theta_{12}=0.323$. The ``\textbf{CKM-like}'' charged lepton corrections are employed, where $\theta_{12}^{l}$ is fixed by Table\,\ref{table1}.}
\end{center}
\end{table}

%% file: prdinput4.tex
\begin{table*}
\begin{center}
\setlength{\tabcolsep}{1.2 em}
\begin{tabular}{l|cccccccc}
\hline
\hline
 $\beta$  & $\theta _{13}^{\nu }\,({}^0)$ & $\theta _{13}^{\nu }\,({}^0)$ & $\frac{\delta }{\pi }$ & $\frac{\delta }{\pi }$ & $\theta _{12}\,({}^0)$ & $\theta _{12}\,({}^0)$ & $\theta _{23}\,({}^0)$ & $\theta _{23}\,({}^0)$ \\
 \hline
\hline
 0.527075 & 7.67 & 7.48 & -0.1681 & -0.1642 & 40.27\,($-$) & 39.88\,($-$) & 47.22\,($1\sigma$)  & 42.46\,($1\sigma$)  \\
 \hline
 0.785059 & 8.32 & -7.92 & 0.4381 & 0.7605 & 37.96 \,($-$)& 37.37 \,($3\sigma$)& 47.31\,($1\sigma$)  & 42.52 \,($1\sigma$) \\
 \hline
 0.952475 & -9.1 & -8.56 & 0.1032 & 0.1068 & 36.46 \,($2\sigma$)& 35.74\,($2\sigma$) & 47.43\,($1\sigma$)  & 42.61\,($1\sigma$)  \\
 \hline
 1 & -9.36 & -8.78 & 0.099 & 0.1024 & 36.04\,($2\sigma$) & 35.28\,($1\sigma$)  & 47.47\,($1\sigma$)  & 42.64\,($1\sigma$)  \\
 \hline
 1.03452 & -9.57 & -8.96 & 0.0961 & 0.0994 & 35.73\,($2\sigma$) & 34.95\,($1\sigma$) & 47.5\,($1\sigma$)  & 42.67 \,($1\sigma$) \\
 \hline
 1.07894 & -9.84 & -9.19 & 0.0925 & 0.09571 & 35.33\,($1\sigma$) & 34.51\,($1\sigma$) & 47.55\,($1\sigma$)  & 42.71\,($1\sigma$)  \\
 \hline
 1.27392 & -11.19 & -10.38 & 0.079 & 0.08141 & 33.57\,($2\sigma$) & 32.61 \,($3\sigma$)& 47.8\,($1\sigma$)  & 42.91\,($1\sigma$)  \\
 \hline
\end{tabular}
\caption{\label{table4}\footnotesize The Scheme-III is tested along with the corrections introduced from different ``\textbf{close to CKM}'' like $U_{lL}$s [see Table.\ref{table1}], with inputs; $\mathbf{\sin^2\theta_{13}\approx 0.023}$, $\mathbf{\delta_{CP}\approx 1.34\,\pi}$ (\textbf{central values}). The predictions of $\theta_{13}^{\nu}$, $\delta$\,(The internal CP phase of $U_{\nu}$), the observable parameters $\theta_{12}$ and $\theta_{23}$ are made in pair. The first column of each pair corresponds to $\phi_{23}=0$ and the rest follows $\phi_{23}=\pi$. This table is associated with $\mathbf{\theta_{13}^{\nu}\lesssim \theta_{C}}$. }
\end{center}
\end{table*}

%% file: prdinput5.tex
\begin{table*}
\begin{center}
\setlength{\tabcolsep}{1.2 em}
\begin{tabular}{l|cccccccc}
\hline
\hline
 $\beta$  & $\theta _{13}^{\nu }\,({}^0)$ & $\theta _{13}^{\nu }\,({}^0)$ & $\frac{\delta }{\pi }$ & $\frac{\delta }{\pi }$ & $\theta _{12}\,({}^0)$ & $\theta _{12}\,({}^0)$ & $\theta _{23}\,({}^0)$ & $\theta _{23}\,({}^0)$ \\
 \hline
\hline
 1 & -15.41 & -15.06 & -0.0518 & -0.0555 & 36.04\,$(2\sigma)$ & 35.28\,$(1\sigma)$ & 48.78\,$(1\sigma)$ & 43.95 \,$(1\sigma)$\\
 \hline
 1.03452 & -15.69 & -15.32 & -0.0508 & -0.0545 & 35.73\,$(2\sigma)$ & 34.95\,$(1\sigma)$ & 48.85\,$(1\sigma)$ & 44.02 \,$(1\sigma)$\\
 \hline
 1.07894 & -16.05 & -15.67 & -0.0495 & -0.0532 & 35.33\,$(1\sigma)$ & 34.51\,$(1\sigma)$ & 48.96 \,$(1\sigma)$& 44.11 \,$(1\sigma)$\\
 \hline
 1.27392 & -17.69 & -17.21 & -0.0443 & -0.0481 & 33.57 \,$(2\sigma)$& 32.61\,$(3\sigma)$ & 49.44\,$(1\sigma)$ & 44.55\,$(1\sigma)$ \\
 \hline
\end{tabular}
\caption{\label{table5}\footnotesize The same description as that of Table.\ref{table4}, but highlights the scenarios when $\mathbf{\theta_{13}^{\nu}\gtrsim \theta_{C}}$. One sees that for $\beta<1$, the $\mathbf{\theta_{13}^{\nu}\gtrsim \theta_{C}}$ predictions are unfounded.}
\end{center}
\end{table*}

%% file: prdinput6.tex
\begin{table*}
\begin{center}
\setlength{\tabcolsep}{1.2 em}
\begin{tabular}{l|cccccccc}
\hline
\hline
 $\beta$  & $\theta _{13}^{\nu }\,({}^0)$ & $\theta _{13}^{\nu}\,({}^0)$ & $\frac{\delta }{\pi}$ & $\frac{\delta }{\pi }$ & $\theta _{12}\,({}^0)$ & $\theta _{12}\,({}^0)$ & $\theta _{23}\,({}^0)$ & $\theta _{23}\,({}^0)$ \\
\hline
\hline
 0.527075 & -9.71 & -9.51 & -0.5247 & -0.521 & 40.26\,$(-)$ & 39.86\,$(-)$ & 47.53\,$(1\sigma)$ & 42.76 \,$(1\sigma)$ \\
\hline
 0.785059 & -11.05 & -10.66 & 0.0915 & 0.0956 & 37.96\,$(-)$ & 37.37\,$(3\sigma)$ & 47.77 \,$(1\sigma)$& 42.96\,$(1\sigma)$ \\
\hline
 0.952475 & -12.86 & -11.6 & -0.0747 & 0.085 & 36.46\,$(2\sigma)$ & 35.74\,$(2\sigma)$ & 48.15\,$(1\sigma)$ & 43.14\,$(1\sigma)$ \\
\hline
 1 & -12.45 & -11.89 & 0.0782 & 0.0823 & 36.04\,$(2\sigma)$  & 35.28\,$(1\sigma)$  & 48.06\,$(1\sigma)$ & 43.2 \,$(1\sigma)$ \\
\hline 
 1.03452 & -12.69 & -12.11 & 0.0763 & 0.0804 & 35.73\,$(2\sigma)$ & 34.95\,$(1\sigma)$ & 48.11\,$(1\sigma)$ & 43.25\,$(1\sigma)$ \\
\hline
 1.07894 & -13.01 & -12.4 & 0.074 & 0.078 & 35.33\,$(1\sigma)$ & 34.51\,$(1\sigma)$ & 48.18\,$(1\sigma)$ & 43.31\,$(1\sigma)$ \\
\hline 
 1.27392 & -14.51 & -13.74 & 0.065 & 0.0688 & 33.57 \,$(2\sigma)$& 32.61\,$(2\sigma)$ & 48.54\,$(1\sigma)$ & 43.62\,$(1\sigma)$ \\
\hline
\end{tabular}
\caption{\label{table6}\footnotesize The depiction of Scheme-III  along with the corrections introduced from ``\textbf{close to CKM}'' like $U_{lL}$s [see Table.\ref{table1}], with inputs; $\mathbf{\sin^2\theta_{13}\approx 0.023}$, $\mathbf{\delta_{CP}\approx 1.48\,\pi}$ (\textbf{central values}). With these, we predict $\theta_{13}^{\nu}$, $\delta$\,(The internal CP phase of $U_{\nu}$), the observable parameters $\theta_{12}$ and $\theta_{23}$. Prediction of each parameter appears in two columns. The first column of corresponds to $\phi_{23}=0$ and the rest is applicable to $\phi_{23}=\pi$. This table contains all the scenarios $\mathbf{\theta_{13}^{\nu}\lesssim \theta_{C}}$ and $\mathbf{\theta_{13}^{\nu}\gtrsim \theta_{C}}$.}
\end{center}
\end{table*}

%% file: prdinput7.tex
\begin{table*}
\begin{center}
\setlength{\tabcolsep}{1.34 em}
\begin{tabular}{l|cccccccc}
\hline
\hline
 $\beta =\alpha$  & $\theta _{13}^{\nu }$ & $\frac{\delta }{\pi }$ & $\frac{\delta _{\text{CP}}}{\pi }$ & $\frac{\delta _{\text{CP}}}{\pi }$ & $\theta _{12}$ & $\theta _{12}$ & $\theta _{23}$ & $\theta _{23}$ \\
 \hline
\hline
 0.527075 & 6.8 & 0.149 & 1.72 & 1.71 & 40.26\,$(-)$ & 39.86\,$(-)$ & 47.11\,$(1\sigma)$ & 42.37\,$(1\sigma)$ \\
 \hline
 0.785059 & 10.14 & 0.217 & 1.57 & 1.55 & 37.96\,$(-)$ & 37.37\,$(3\sigma)$ & 47.6\,$(1\sigma)$ & 42.87 \,$(1\sigma)$\\
 \hline
 0.952475 & 12.3 & 0.2388 & 1.51 & 1.48 & 36.46\,$(3\sigma)$ & 35.74\,$(1\sigma)$ & 48.03\,$(1\sigma)$ & 43.29 \,$(1\sigma)$\\
  \hline
 1 & 12.92 & 0.2436 & 1.49 & 1.47 & 36.04\,$(2\sigma)$ & 35.28\,$(1\sigma)$ & 48.16\,$(1\sigma)$ & 43.42\,$(1\sigma)$ \\
  \hline
 1.03452 & 13.36 & 0.2469 & 1.48 & 1.46 & 35.73\,$(1\sigma)$ & 34.95\,$(1\sigma)$ & 48.26\,$(1\sigma)$ & 43.53\,$(1\sigma)$ \\
  \hline
 1.07894 & 13.94 & 0.25 & 1.47 & 1.44 & 35.33\,$(1\sigma)$ & 34.51 \,$(1\sigma)$& 48.4\,$(1\sigma)$ & 43.66\,$(1\sigma)$ \\
  \hline
 1.27392 & 16.45 & 0.2647 & 1.42 & 1.38 & 33.57\,$(2\sigma)$ & 32.61 \,$(3\sigma)$& 49.07\,$(1\sigma)$ & 44.33\,$(1\sigma)$ \\
  \hline
\end{tabular}
\caption{\label{table7}\footnotesize In \textbf{Scheme-III}, one sees $\theta_{13}^{\nu}\simeq\theta_{C}$, similar to the $1$-$2$ rotation angle, $\theta_{12}^{l}$ which is also,   $\theta_{12}^{l}\simeq\theta_{C}$. This motivates one to look into those possibilities, where, $\theta_{13}^{\nu}=\theta_{12}^{l}\simeq \theta_{C}$; which says $\alpha=\beta$. It is found that this possibilities are more relevant for the cases, $\beta \geq 1$. }
\end{center}
\end{table*}